\documentclass{emulateapj}  
\usepackage{graphicx} 
\usepackage{apjfonts}

\slugcomment{To appear in ApJ Letters}
\shorttitle{Spectropolarimetry of dark-cored penumbral filaments} 
\shortauthors{Bellot Rubio et al.}
\begin{document}

\title{Vector spectropolarimetry of dark-cored penumbral filaments with
{\em Hinode}} 

\author{L.R.\ Bellot Rubio\altaffilmark{1}, S.\ Tsuneta\altaffilmark{2}, K.\
Ichimoto\altaffilmark{2}, Y.\ Katsukawa\altaffilmark{2}, B.W.\
Lites\altaffilmark{3}, \\ S.\ Nagata\altaffilmark{4}, T.\
Shimizu\altaffilmark{5}, R.A.\ Shine\altaffilmark{6}, Y.\
Suematsu\altaffilmark{2}, T.D.\ Tarbell\altaffilmark{6}, \\ A.M.\
Title\altaffilmark{6}, and J.C.\ del Toro Iniesta\altaffilmark{1} }

\altaffiltext{1}{Instituto de Astrof\'{\i}sica de
Andaluc\'{\i}a (CSIC), Apdo.\ de Correos 3004, 18080 Granada, Spain} 

\altaffiltext{2}{National Astronomical Observatory of Japan,
2-21-1 Osawa, Mitaka, Tokyo 181-8588, Japan} 

\altaffiltext{3}{High Altitude
Observatory, NCAR, 3080 Center Green Dr.\ CG-1, Boulder, CO 80301, USA}

\altaffiltext{4}{Hida Observatory, Kyoto University, Takayama, Gifu 506-1314,
Japan} 

\altaffiltext{5}{Institute of Space and Astronautical Science, JAXA,
Sagamihara, Kanagawa 229-8510, Japan} 

\altaffiltext{6}{Lockheed Martin Solar
and Astrophysics Laboratory, Bldg.\ 252, 3251 Hanover St., Palo Alto, CA
94304, USA}

\begin{abstract} 
We present spectropolarimetric measurements of dark-cored penumbral filaments
taken with Hinode at a resolution of 0\farcs3. Our observations demonstrate
that dark-cored filaments are more prominent in polarized light than in
continuum intensity. Far from disk center, the Stokes profiles emerging from
these structures are very asymmetric and show evidence for magnetic fields of
different inclinations along the line of sight, together with strong Evershed
flows of at least 6-7 km~s$^{-1}$.  In sunspots closer to disk center,
dark-cored penumbral filaments exhibit regular Stokes profiles with little
asymmetries due to the vanishing line-of-sight component of the
horizontal Evershed flow. An inversion of the observed spectra indicates that
the magnetic field is weaker and more inclined in the dark cores as compared
with the surrounding bright structures. This is compatible with the idea that
dark-cored filaments are the manifestation of flux tubes carrying hot Evershed
flows.
\end{abstract}


\section{Introduction}
\label{sec:intro}

Since their discovery by Scharmer et al.\ (2002), dark-cored penumbral
filaments have attracted much interest because they could reveal the primary
mode of magnetoconvection taking place in sunspot penumbrae.  The dark cores
of penumbral filaments are narrow ($\leq 200$~km) lanes running along the
filament axes, surrounded by two lateral brightenings. It has been proposed
that dark-cored penumbral filaments are the manifestation of magnetic flux
tubes (S\"utterlin et al.\ 2004; Bellot Rubio et al.\ 2005; Borrero 2007) or
field-free gaps in the penumbra (Spruit \& Scharmer 2006; Scharmer \& Spruit
2006), but no consensus has been reached yet.

The morphology and temporal evolution of these structures are well known from
high-resolution (0\farcs1--0\farcs2) filtergrams.  Imaging observations show
the two lateral brightenings to move with the same speed and direction as a
single entity. Spectroscopic measurements at 0\farcs2 resolution (Bellot Rubio
et al.\ 2005; Rimmele \& Marino 2006) have revealed that the Evershed flow is
concentrated preferentially in the dark cores, where it seems to be an upflow
near the umbra/penumbra boundary. They also provide some evidence that the
dark cores possess weaker magnetic fields than their lateral brightenings.
One-wavelength longitudinal magnetograms at 0\farcs2 resolution (Langhans et
al.\ 2005; Langhans et al.\ 2007) show lower circular polarization signals in
the dark cores as compared with the bright edges, supporting the conclusions
derived from spectroscopy. However, high-precision vector magnetic field
measurements of these structures have never been made because of the
relatively modest angular resolution attained by ground-based
spectropolarimeters.

In this Letter, we report on the first spectropolarimetric measurements of
dark-cored penumbral filaments. The observations were taken with Hinode's
Solar Optical Telescope at a resolution of 0\farcs3. Dark-cored filaments turn
out to be prominent structures in polarized light, showing anomalous, very
asymmetric Stokes $V$ profiles in sunspots outside the disk center. In
addition, we find that the dark cores possess weaker and more horizontal
fields than their surroundings.

\section{Observations}
\label{sec:obs}

We used the spectropolarimeter (SP; Lites et al.\ 2001) of the Solar Optical
Telescope (SOT) aboard Hinode (Kosugi et al.\ 2007) to observe the \ion{Fe}{1}
630.2~nm lines emerging from AR 10923 on November 10 and 14, 2006. On those
days the spot was located at heliocentric angles of $50^\circ$ and $8^\circ$,
respectively. The four Stokes profiles of the two \ion{Fe}{1} lines were
measured by scanning the spectrograph slit across the spot in steps of
0\farcs148, with a wavelength sampling of 2.153~pm~pixel$^{-1}$.  The SP slit
width and the CCD pixel size are 0\farcs16. The integration time per slit
position was 4.8~s, resulting in a noise level of $10^{-3} I_{\rm c}$ as
measured in the continuum of Stokes $Q$, $U$ and $V$. Image stabilization down
to 0\farcs01 rms was provided by the SOT tip-tilt system (Shimizu et al.\
2007). These measurements reach a spatial resolution of about 0\farcs32, while
the diffraction limit of the SOT is $\lambda/D = 0\farcs25$ at 630.2~nm.

The data have been corrected for dark current, flat field, and instrumental
polarization. We estimate that residual instrumental crosstalk is not larger
than a few $10^{-4} I_{\rm c}$. The calibrated Stokes profiles have been used
to create maps of continuum intensity, $I_{\rm c}$, total circular
polarization, ${\rm TCP} = \int |V| \, {\rm d}\lambda$, total linear
polarization, ${\rm TLP} = \int (Q^2+U^2)^{1/2} \, {\rm d}\lambda$, total
polarization, ${\rm TP} = \int (Q^2+U^2+V^2)^{1/2} \, {\rm d}\lambda$, and
area asymmetry, ${\delta A} = \int V \, {\rm d}\lambda/ \int |V| \, {\rm
d}\lambda$, where the integrals extend over a wavelength range of 0.1098~nm to
encompass the 630.25~nm line.

\section{General appearance and Stokes spectra}
\begin{figure}
\centering
\resizebox{1.01\hsize}{!}{\includegraphics{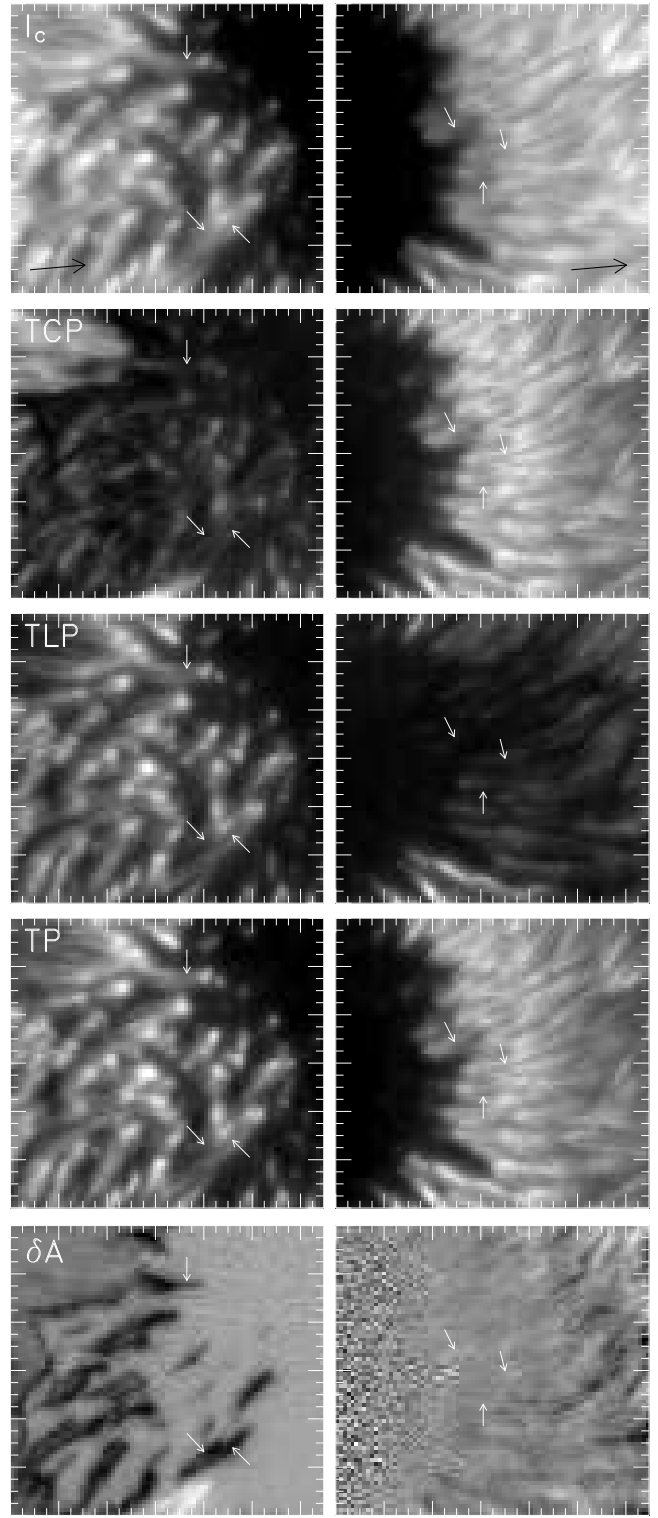} }
\caption{Dark-cored penumbral filaments observed on the limb side (left) and
center side (right) of AR 10923 on November 10, 2006, between 16:01 and 17:26
UT. The heliocentric angle of the spot was $50^\circ$. North is up and West to
the right. Minor tick marks are separated by 0\farcs5. The black arrows point
to disk center.  From top to bottom: maps of continuum intensity, total
circular polarization, total linear polarization, total polarization, and area
asymmetry for the two regions. White arrows indicate the position of
dark-cored filaments.}
\label{fig1}
\end{figure}

Figure~\ref{fig1} shows two $12\farcs8 \times 12\arcsec$ regions in the
limb-side and center-side penumbra of AR 10923 at a heliocentric angle of
50$^\circ$. The small white arrows mark the positions of six dark-cored
penumbral filaments. As noticed by S\"utterlin et al.\ (2004), the dark cores
show higher continuum intensity contrasts in the center-side penumbra (right
panels).

We find the dark-cored filaments to be much more prominent in polarized light
than in continuum intensity (compare, e.g., the first and fourth panels of
Figure~\ref{fig1}). This is also the case for the November 14 observations,
when the spot was at only $8^\circ$ from disk center.  The fact that
dark-cored filaments show up more prominently in polarized light implies that
there is no one-to-one correspondence between the underlying magnetic fields
and their continuum intensity signatures. Thus, a complete characterization 
of the penumbra cannot be based solely on intensity measurements.

The dark cores exhibit smaller circular and linear polarization signals than
the bright structures surrounding them, which is highly suggestive of weaker
fields. Contrary to one-wavelength magnetograms, the TCP, TLP and TP signals
are not affected by velocities, but no unambiguous conclusion on field
strengths can be drawn from simple inspections of these parameters because
they still depend on temperature.

An interesting fact is that the dark-cored filaments are detected as
structures of enhanced Stokes $V$ area asymmetries in the limb-side penumbra
of spots away from disk center (bottom left panel of Fig.~\ref{fig1}). On the
center side they are not distinct features in area asymmetry. Very likely,
this is the result of projection effects that shift the optical depth scale
toward higher geometrical heights and thus place the Evershed flow deeper in
the line forming region or even outside of it. The enhanced area asymmetries
imply the existence of gradients/discontinuities of velocity (and possibly
other atmospheric parameters) along the line of sight (LOS) which are 
stronger in the filaments than in the surroundings.

\begin{figure*}[t]
\centering \resizebox{.19\hsize}{!}{\includegraphics[bb=107 352 326
540]{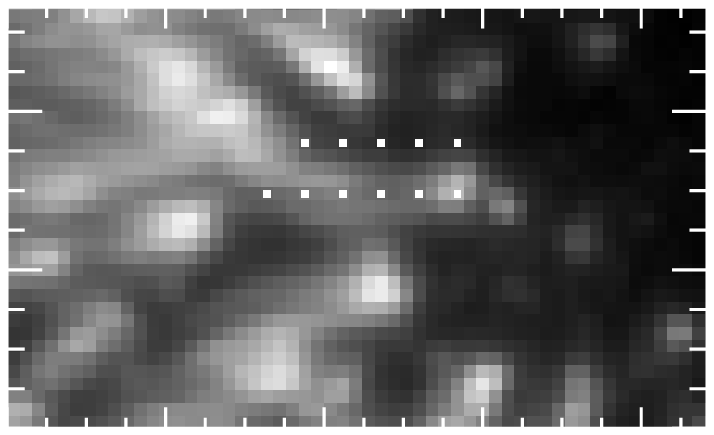}} 
\resizebox{.8\hsize}{!}{\includegraphics[bb=60 365 702
515]{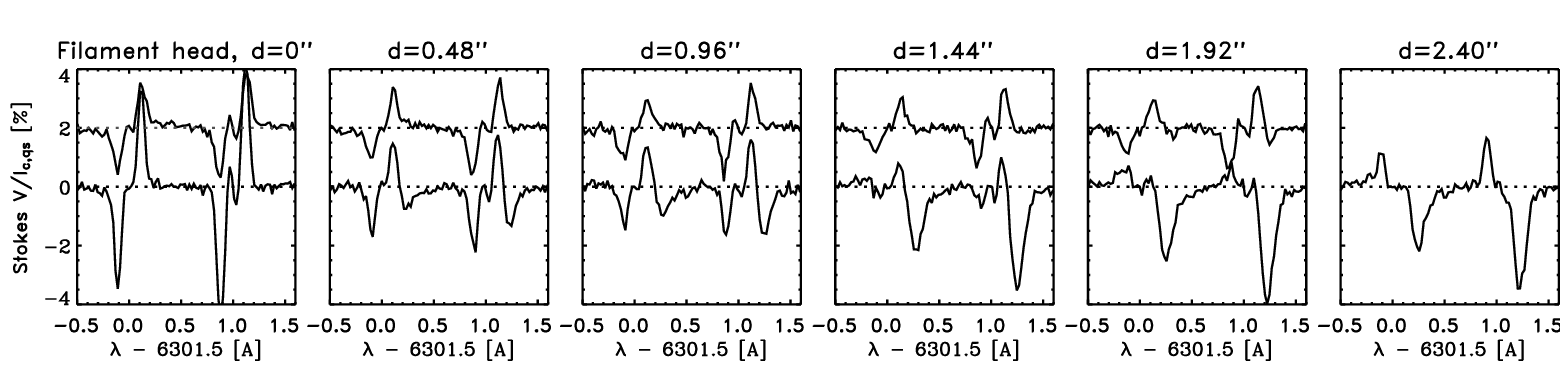} }
\caption{\ion{Fe}{1}~630.15 and 630.25~nm Stokes $V$ profiles emerging from a
dark-cored filament in the limb-side penumbra and the surrounding umbra as a
function of distance to the filament head. The distance increases in steps of
0\farcs48 from right to left (total polarization map) and from left to right
(Stokes spectra panels). The umbra is sampled 0\farcs64 away from the dark
core; the corresponding Stokes profiles are shifted upward by 2\% for better
visibility. The dark-cored filament makes an angle of $7^\circ$ to the line of
symmetry. The polarity of the spot is negative.}
\label{fig2}
\end{figure*}

\begin{figure}
\centering
\resizebox{.93\hsize}{!}{\includegraphics[bb=20 60 450 690]
{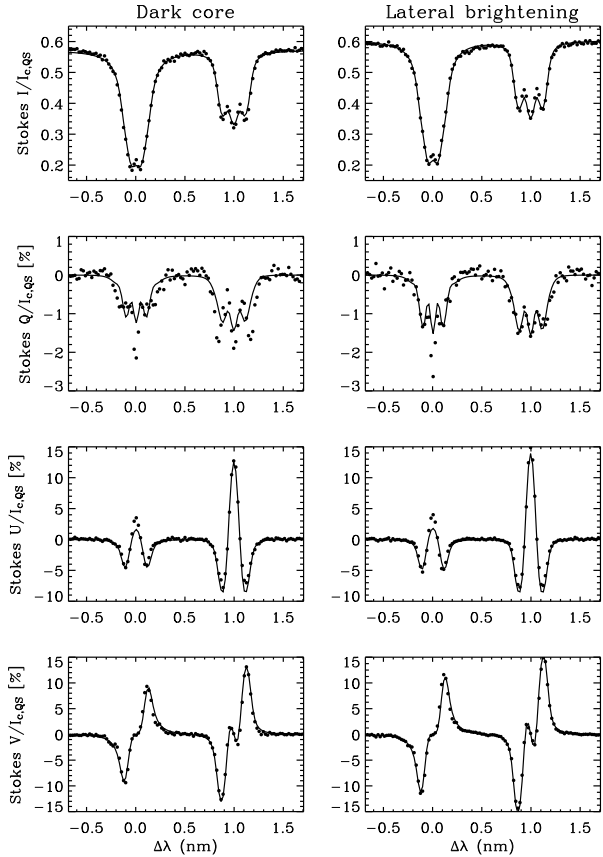}}
\caption{Observed (dots) and best-fit (solid lines) profiles from a
one-component inversion with SIR, for a pixel inside a dark core (left) and
its lateral brightening (right).  Both structures were observed in the
limb-side penumbra of AR 10923 at $8^\circ$ from disk center
(Fig.~\ref{fig4}).}
\label{fig3}
\end{figure}

This conclusion is supported by the anomalous Stokes $V$ spectra emerging from
dark-cored filaments. Figure~\ref{fig2} displays the circular polarization
profiles observed along the dark core of a limb-side filament near the inner
penumbral boundary.  For comparison, we also plot the spectra from the 
surrounding umbra at the same radial distances (cf.\ the dots in the left
panel of Fig.~\ref{fig2}). The bright penumbral grain that forms the head of
the filament exhibits regular Stokes $V$ profiles with the polarity of the
spot. Further outward along the dark core, the circular polarization profiles 
show three lobes indicating the existence of opposite polarity fields along 
the LOS, {\em one of which is strongly Doppler-shifted to the red}. Its 
LOS velocity can roughly be estimated as $6-7$ km~s$^{-1}$ from the wavelength 
separation between the redmost Stokes $V$ peaks.  This magnetic field component 
gradually takes over, giving rise to one-lobed Stokes $V$ spectra at a distance 
of $d \sim 1\farcs92$ and two-lobed profiles of opposite polarity farther than 
$d = 2\farcs4$.  In contrast, the circular polarization profiles emerging from 
the adjacent background umbra keep the polarity of the spot at all radial 
distances, and do not exhibit anomalous shapes.

The multi-lobed Stokes $V$ spectra observed in limb-side filaments at 0\farcs3
resemble those measured from the ground at much lower angular resolution.  In
our data, we can for the first time distinguish between the dark cores and the
lateral brightenings, each of which may differ in their magnetic properties. 
Thus, these measurements constrain horizontal variations of the physical
parameters relative to those along the LOS much more tightly than earlier 
observations. Both Fig.~2 and the observed Stokes $V$ asymmetries indicate a
complex magnetic topology in the dark-cored filaments, with discontinuities or
gradients along the LOS, and possibly also lateral gradients.  To explain the 
observations, the height variations/discontinuities of atmospheric parameters 
must be present well within the Stokes $V$ formation region (i.e., significantly 
above the $\tau = 1$ level; Cabrera Solana et al.\ 2005), as otherwise they would 
be unable to produce the multi-lobed profiles depicted in Fig.~\ref{fig2}. 
Finally, we note that one-wavelength longitudinal magnetograms of complex 
spectra such as the ones discussed here may even fail in retrieving the 
correct polarity of the field, depending on the exact wavelength of the 
observations.

\section{Magnetic fields of dark-cored filaments}
We estimate the vector magnetic field of dark-cored penumbral filaments and
their surroundings by inverting the observed spectra with the SIR code (Ruiz
Cobo \& del Toro Iniesta 1992). We restrict our analysis to the November 14
data set, since on that day the Stokes profiles show negligible asymmetries in
the absence of strong LOS velocities (due to the proximity of the spot to disk
center). The spectra can be fitted remarkably well in terms of one-component
model atmospheres featuring constant (height-independent) vector magnetic
fields, LOS velocities, and microturbulent velocities (see Fig.~\ref{fig3}).
The temperature stratification is determined by changing the slope and
$T(\tau=1)$ value of the penumbral model of Bellot Rubio et al.\ (2006).  In
total, the inversion returns the values of 8 free parameters. We assume zero
stray light contamination and unity magnetic filling factors.

\begin{figure*}[t]
\centering
\resizebox{.96\hsize}{!}{
\includegraphics[bb=105 395 517 620]{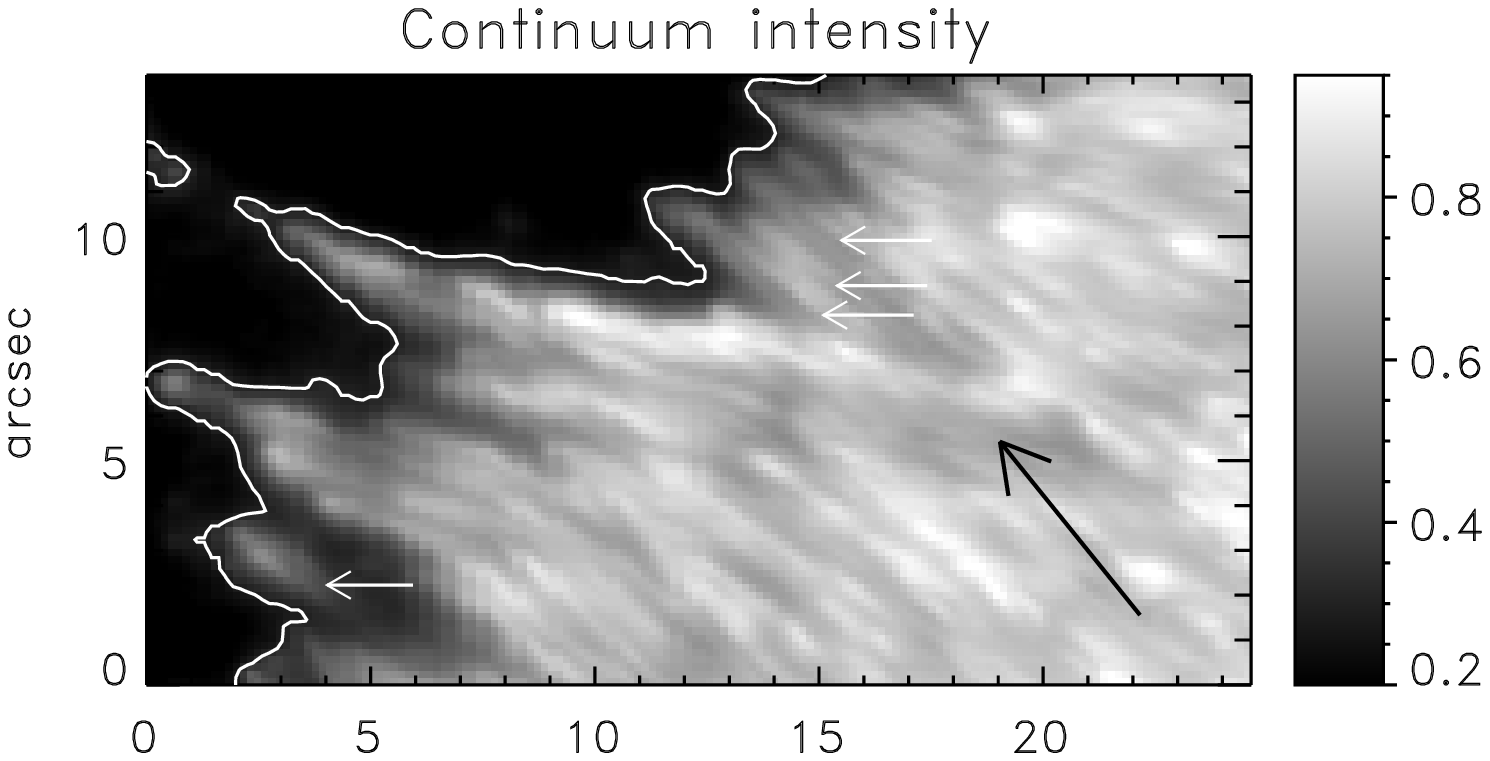}
\includegraphics[bb=105 395 517 620]{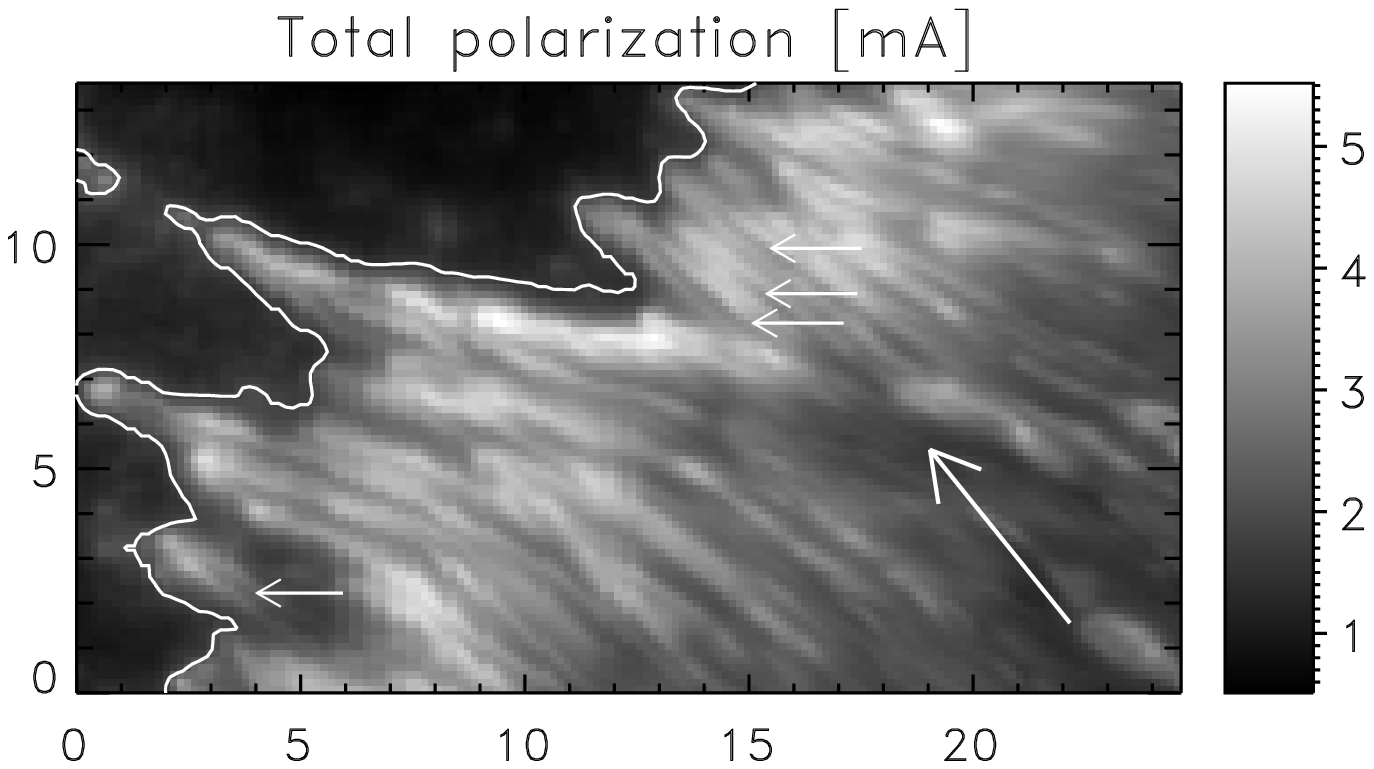}} 
\resizebox{.96\hsize}{!}{
\includegraphics[bb=104 395 516 620]{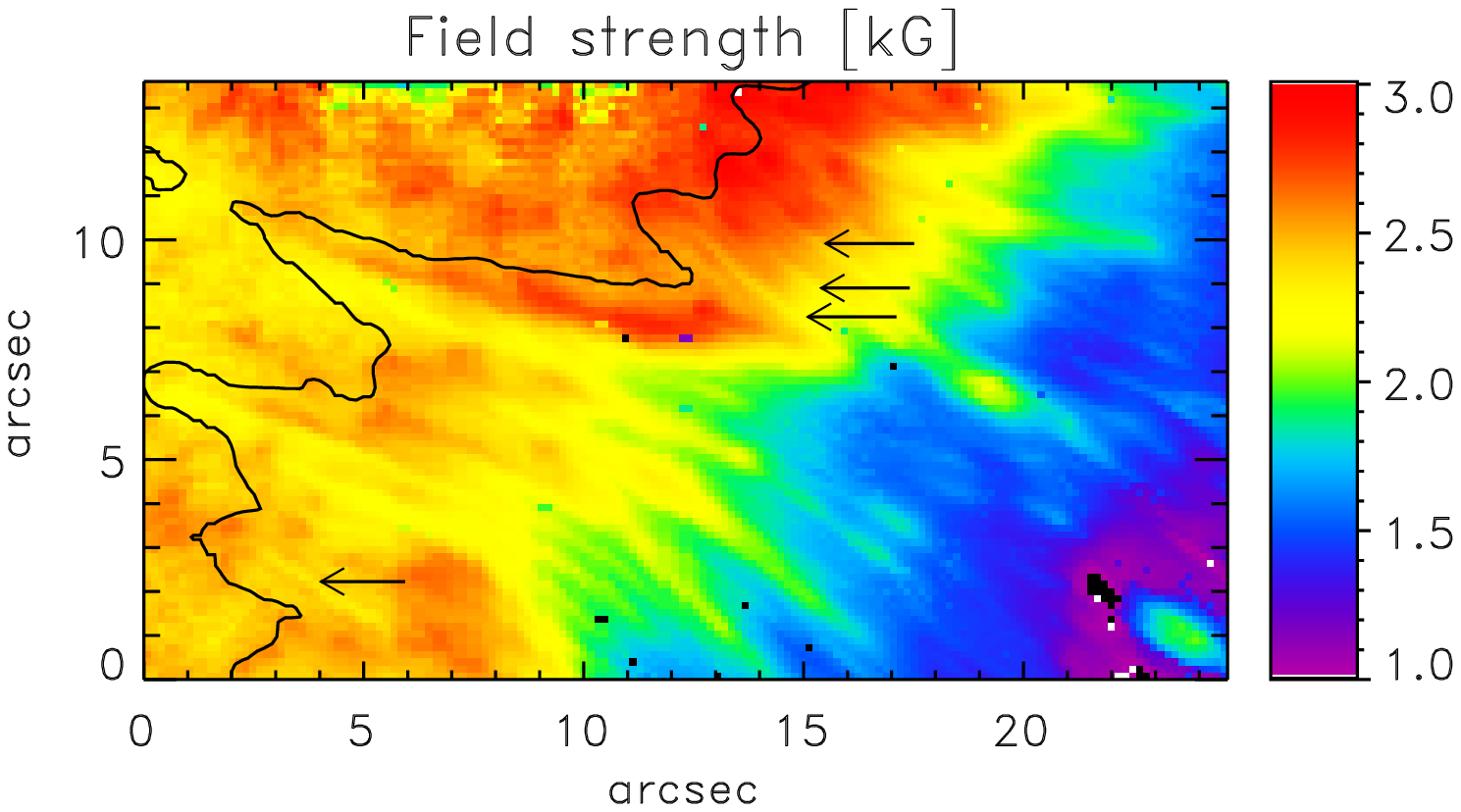}
\includegraphics[bb=103 395 515 620]{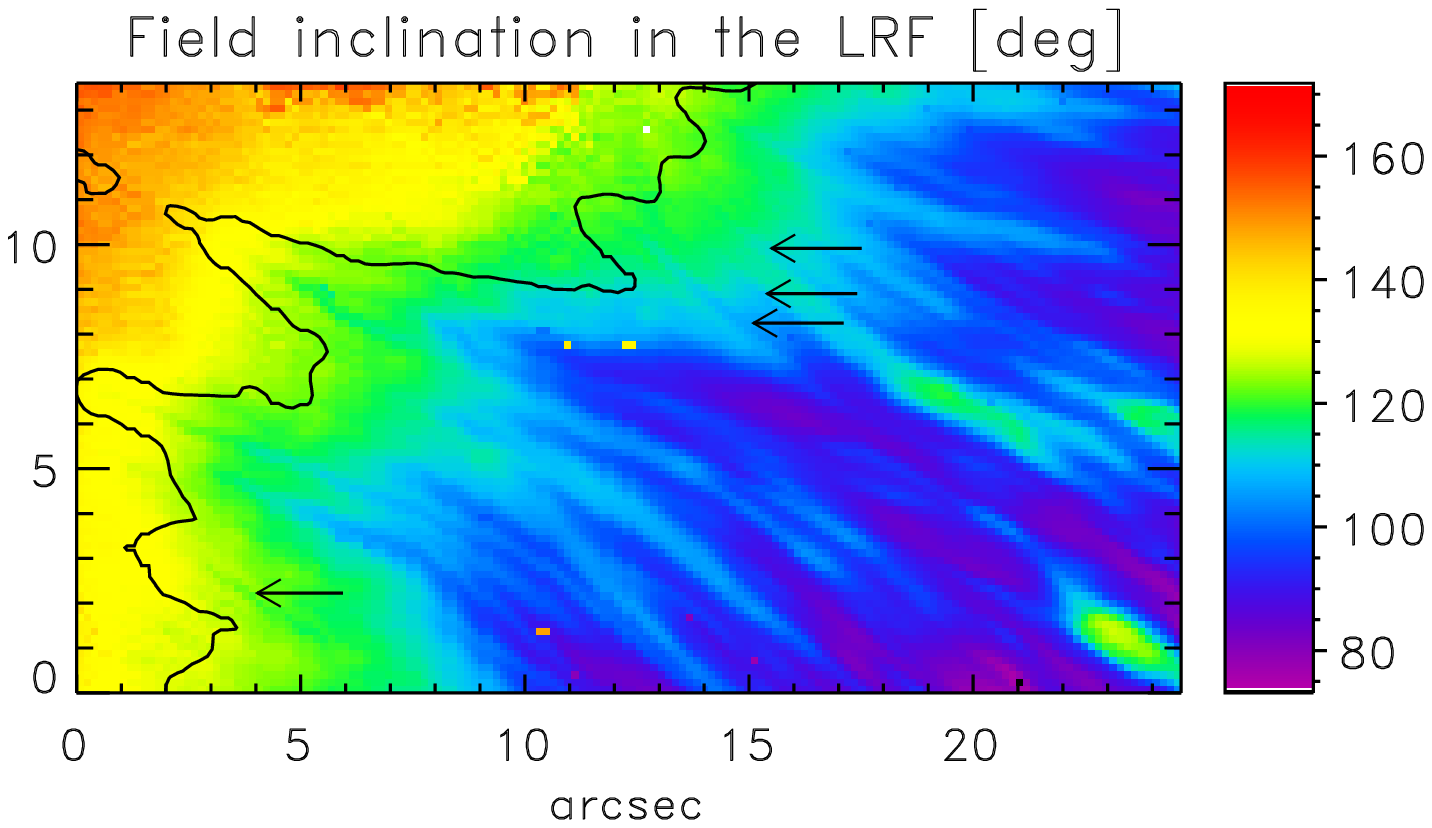}}
\vspace*{1.5em}
\caption{{\em Top:} Continuum intensity (left) and total polarization
(right) maps of the limb-side penumbra of AR 10923 as observed by
Hinode/SP on November 14, 2006 between 16:30 and 17:15 UT. Four
dark-cored filaments are indicated with small horizontal arrows. The
long arrows point to disk center.  {\em Bottom:} Maps of
magnetic field strength and inclination in the LRF, as
deduced from an inversion of the spectra. }
\vspace*{.5em}
\label{fig4}
\end{figure*}

Figure~\ref{fig4} displays continuum intensity and total polarization maps of
a small region in the limb-side penumbra of AR 10923 on November 14. At least
four dark-cored filaments can easily be recognized, especially in the
polarization map. Also shown in Fig.~\ref{fig4} are the field strengths and
inclinations resulting from the inversion, expressed in the local reference
frame (LRF).  The dark cores stand out clearly as long, narrow streaks of
weaker ($\Delta B \sim 100-150$ G) and more inclined ($\Delta \gamma_{\rm LRF}
\sim 4^\circ$) fields than the lateral brightenings or surrounding structures.
In general, we find a very smooth magnetic topology in this part of the
penumbra, which is only altered by the presence of the dark cores. Note that
the inclination map shows extended patches in the penumbra where the polarity
of the magnetic field is opposite to that of the spot, i.e., field lines
returning back to the solar surface ($\gamma_{\rm LRF} <90^{\rm o}$).  The
existence of such field lines has been inferred from inversions of Stokes
profiles at lower resolution, but this is the first time they are detected so
close to the umbra (see also Ichimoto et al.\ 2007). As pointed out by Bellot
Rubio (2007), the existence of field lines returning to the solar surface well
within the penumbra may pose a serious problem for the gappy model of Spruit
\& Scharmer (2006).

%

\section{Conclusions}
\label{sec:con}

The nearly diffraction-limited (0\farcs3) spectropolarimetric measurements
performed by Hinode reveal that dark-cored penumbral filaments are conspicuous
structures in polarized light, showing higher contrasts than in continuum
intensity. The total polarization signal emerging from the dark cores is
smaller than that observed in the lateral brightenings, which is highly
suggestive of lower field strengths. We have confirmed this hyphotesis by
carrying out an inversion of the spectra observed in a sunspot very close 
to disk center.

The dark cores possess $100-150$~G weaker fields than the lateral brightenings
and surrounding structures, i.e., they are not field-free. In addition, the
field is deduced to be more horizontal in the dark cores than in the lateral
brightenings, by some $4^\circ$. The field strength and inclination
differences are surprisingly small, which might indicate still insufficient
angular resolution and/or simplistic modeling (the inversions performed here
just provide {\em average} values of the atmospheric parameters along the
LOS). High-resolution magnetograph observations (Langhans et al.\ 2007)
suggest that the actual differences could be larger.

Our results are consistent with the idea that dark-cored filaments represent
magnetic flux tubes embedded in an ambient field. Their properties coincide
with those inferred from uncombed inversions of visible and infrared lines
(weaker and more inclined fields in the tubes as compared with the background
atmosphere; large area asymmetries and multi-lobed Stokes $V$ profiles caused
by strong Evershed flows along the tubes; see, e.g., Beck 2006 and Borrero et
al.\ 2006). They also agree with the properties of moving penumbral tubes as
deduced from numerical simulations (Schlichenmaier et al.\ 1998;
Schlichenmaier 2002). According to the simulations, the field is weaker in the
tubes to maintain horizontal mechanical equilibrium. It is also more inclined
because the tubes rapidly bend when they reach the subadiabatic layers of the
photosphere during their ascent. In the moving tube simulations, Evershed
flows exceding 10 km~s$^{-1}$ are driven by gas pressure gradients that build
up as a consequence of the radiative cooling of the plasma. By contrast, no
mechanism capable of explaining Evershed flows of several km~s$^{-1}$ has been
identified in the gappy penumbral model as yet. The Eveshed flow is an
ingredient without which the observed Stokes $V$ shapes and area asymmetries
cannot be understood.

Hopefully, further analyses of Hinode spectropolarimetric observations will
help distinguish between competing models of the penumbra (Bellot Rubio
2007). Initial efforts in this direction are presented by Jur\v{c}ak et
al.\ (2007) and Ichimoto et al.\ (2007).

\acknowledgments 

Hinode is a Japanese mission developed and launched by ISAS/JAXA, with NAOJ as
domestic partner and NASA and STFC (UK) as international partners. It is
operated by these agencies in co-operation with ESA and NSC (Norway). This
work has been partially funded by the Spanish MEC through project
ESP2006-13030-C06-02.

\end{document}